\documentstyle[fleqn,12pt,english]{article}
\def\beq{\begin{equation}}
\def\eeq{\end{equation}}
\def\0{\otimes}
\def\1{\mbox{1\hskip-.25em l}}
\def\6{\langle}
\def\9{\rangle}
\def\half{\mbox{$1\over2$}}
\def\bm{\mbox{\boldmath$m$}}
\def\bn{\mbox{\boldmath$n$}}
\def\tx{\theta_x}
\def\tz{\theta_z}
\def\px{\phi_x}
\def\pz{\phi_z}

\begin{document}
\begin{center}
{\large{\bf Covariant quantum measurements may not be optimal}}\\[10mm]
ASHER PERES and PETRA F. SCUDO\\[8mm]
Department of Physics, Technion---Israel Institute of
Technology, 32000 Haifa, Israel\end{center}\vskip15mm

\noindent{\bf Abstract. }\ Quantum particles, such as spins, can be used
for communicating spatial directions to observers who share no common
coordinate frame. We show that if the emitter's signals are the orbit
of a group, then the optimal detection method may not be a covariant
measurement (contrary to widespread belief).  It may be advantageous
for the receiver to use a different group and an indirect estimation
method: first, an ordinary measurement supplies redundant numerical
parameters; the latter are then used for a nonlinear optimal
identification of the signal.
\vskip15mm

\noindent{\bf1. \ Indiscrete quantum information}\footnote{Note to
printer: this is not a typo. The term we use is indiscrete (meaning not 
discrete). Do not confuse that with the word indiscreet.}

Information theory usually deals with the transmission of a sequence
of {\it discrete\/} symbols, such as 0 and 1. Even if the information
to be transmitted is of continuous nature, such as the position of a
particle, it can be represented with arbitrary accuracy by a string of
bits. However, there are situations where information cannot be encoded
in such a way. For example, the emitter (conventionally called Alice)
wants to indicate to the receiver (Bob) a direction in space.  If they
have a common coordinate system to which they can refer, or if they can
create one by observing distant fixed stars, Alice simply communicates
to Bob the components of a unit vector \bn\ along that direction, or its
spherical coordinates $\theta$ and $\phi$. But if no common coordinate
system has been established, all she can do is to send a real physical
object, such as a gyroscope, whose orientation is deemed stable.

In the quantum world, the role of the gyroscope is played by a system
with large spin. Earlier works [1--4] considered the use of spins
for transmitting a single direction. The simplest method~\cite{mp}
is to send these spins polarized along the direction that one wishes
to indicate. This, however, is not the most efficient procedure: when
two spins are transmitted, a higher accuracy is achieved by preparing
them with {\it opposite\/} polarizations~\cite{gp}. If there are more
than two spins, optimal results are obtained with {\it entangled\/}
states \cite{bbbmt,petra}.

The fidelity of the transmission is usually defined as

\beq F=\6\cos^2(\chi/2)\9=(1+\6\cos\chi\9)/2, \eeq
where $\chi$ is the angle between the true \bn\ and the direction
indicated by Bob's measurement. The physical meaning of $F$ is that
the infidelity,

\beq 1-F=\6\sin^2(\chi/2)\9,\eeq
is the mean square error of the measurement, if the error is defined
as $\sin(\chi/2)$~\cite{helstrom}. The experimenter's aim, minimizing
the mean square error, is the same as maximizing fidelity. We can of
course define ``error'' in a different way, and then fidelity becomes
a different function of $\chi$ and optimization leads to different
results. With the definition in Eq.~(\theequation), it can be shown
\cite{bbbmt,petra} that for a large number $N$ of spins, the infidelity
asymptotically tends to

\beq 1-F=5.783/N^2=1.446/d, \label{infi1}\eeq
where $d$ is the dimension of the subspace of Hilbert space
that is effectively used for the transmission.

A more difficult problem is the transmission of a complete Cartesian
frame, if a single quantum messenger is available. In an earlier
publication~\cite{euler}, we showed how a hydrogen atom (formally,
a spinless particle in a Coulomb potential) can transmit a complete
frame. We assumed the hydrogen atom to be in a Rydberg state (an energy
eigenstate is needed to ensure the stability of the transmission). The
$n$-th energy level of that atom has degeneracy $d=n^2$ because the
total angular momentum may take values $j=0,\cdots,n-1$, and for each
one of them $m=-j,\cdots,j$.  A similar calculation was done by Bagan,
Baig, and Mu\~noz-Tapia (hereafter BBM)~\cite{bbm}, who were able
to reach much higher values of $j$ and to prove that the asymptotic
behavior was

\beq 1-F\to 1/\sqrt{d}. \eeq
Here the infidelity $1-F$ is the sum of the mean square errors for
three orthogonal axes.

There is an essential difference between our work and that of BBM. We
considered a single system (a hydrogen atom in a Rydberg state),
while BBM took $N$ spins, and one irreducible representation for each
value $j$ of the total angular momentum. The maximum value is $j_{{\rm
max}}=N/2$, and then the mathematics are the same as for our Rydberg
state, with $j_{{\rm max}}= n-1$, as explained above.  However,
if there are $N$ spins that can be sent independently, there is a
better method. Alice can use half of them to indicate her $z$ axis,
and the other half for her $x$ axis. The two directions found by Bob
may not be exactly perpendicular, because separate transmissions have
independent errors due to limited angular resolution. Some adjustment
will be needed to obtain Bob's best estimates for the $z$ and $x$ axes,
before he can infer from them his guess of Alice's $y$ direction. Even
without this adjustment, this method is far more accurate, especially
if $N$ is large. From Eq.~(\ref{infi1}), the mean square error for
each one of the $z$ and $x$ axes is $5.783/(N/2)^2=23.13/N^2$, rather
than $4/3N$ which is the result with the method used by BBM~\cite{bbm}.

Similar results hold even for low values of $N$. For example, if Alice
has four spins at her disposal, she can do better in this way than with
the BBM method: she sends two spins with opposite polarizations along
her $z$ axis (the Gisin-Popescu method~\cite{gp}) and two with opposite
polarizations along her $x$ axis. The infidelity for each one of these
axes is 0.21132 (this can still be improved by forcing orthogonality
on Bob's axes, as explained in Sect.~4). On the other hand, with a hydrogen
atom~\cite{euler} and $j_{{\rm max}}=2$, if we optimize two axes, the
mean square error per axis is 0.23865. Why is there such a discrepancy?

\bigskip

\noindent{\bf2. \ Covariant measurements are not always optimal}

In all the works that were mentioned above, and in many other similar
ones, it was assumed that Holevo's method of covariant measurements
\cite{holevo} gave optimal results. That method considers the
case where Alice's signals are the orbit of a group $\cal G$, with
elements~$g$. Namely, if $|A\9$ is one of the signals, the others
are $|A_g\9=U(g)|A\9$, where $U(g)$ is a unitary representation of
the group element $g$.  The only problem was to find optimal quantum
states for Alice's signals and Bob's detectors.  Originally, Holevo
considered only irreducible representations. It is known now that in
some cases reducible representations are preferable \cite{bbbmt,petra}.
One then never needs to use more than one copy of each irreducible
representation in the reducible one. For example, if $|A_g\9$ has four
spins as in the above example, this state can be written by using each
one of $j=0,1,2$ only once, as shown explicitly in the next section.

We now turn our attention to Bob.  The mathematical representation
of his apparatus is a {\it positive operator valued measure\/}
(POVM)~\cite{qt}, namely a resolution of identity by a set of positive
operators:

\beq  \sum_h E_h = \1,\eeq
where the label $h$ indicates the outcome of Bob's experiment.  This is
true for {\it any\/} type of measurement, provided that the labels $h$
are kept ``raw'' and not subjected to further classical processing
into a new set of labels, as explained in Sect.~4 and~5. In the case
of {\it covariant\/} measurements, the labels $h$ run over all the elements of
the group $\cal G$ (with a suitable adjustment of the notation in the
case of continuous groups). Then the probability that Bob's apparatus
indicates group element $h$ when Alice sent a signal $|A_g\9$ is

\beq P(h|g)=\6A_g|E_h|A_g\9. \label{Phg}\eeq
The method of covariant measurements further assumes that $E_h$ can
be written as

\beq E_h=|B_h\9\6B_h|,\label{cov} \eeq
where

\beq |B_h\9=U(h)|B\9. \eeq
Here, $|B\9$ is a fiducial vector for Bob (which has to be optimized)
and $U(h)$ is a representation (possibly a direct sum of irreducible
representations) of the same group $\cal G$ that Alice is using.

All this seems quite reasonable (and this indeed usually works well)
but, as the above example of four spins shows, this may not be the
optimal method. In that example, Alice's signals $|A_g\9$, for all
possible positions of her axes, are $SO(3)$ rotations of a fiducial
state $|A\9$ with $j=0,1,2$ (see next section for details). On the
other hand Bob uses two separate POVMs, each one testing only two of
the four spins. Each one of these POVMs also involves $SO(3)$,
but with $j=0$ and~1 only. (Strictly speaking, the relevant
mathematical structure is $S_2\otimes S_2$, where $S_2$ is the
quotient $SO(3)/SO(2)$, namely the two-dimensional sphere which is
not a group. We shall ignore this technical point and informally call
it a group, to avoid unnecessarily cumbersome terminology.)

\bigskip

\noindent{\bf3. \ Equivalent irreducible representations}

Let us examine carefully the meaning and construction of equivalent
irreducible representations.  Unitary equivalence is {\it not\/}
equivalence from the point of view of physics~\cite{fong}. A simple
example is a particle of spin~$3\over2$, whose state space has
four dimensions and is unitarily equivalent to that of a pair of
spin~$1\over2$ particles. In atomic physics, unitarily equivalent
representation naturally arise when we consider different couplings
of the various spins, and we use Clebsch-Gordan coefficients in order to
construct new states in a systematic way. These ``equivalent'' unitary
representations actually correspond to quite different states. For
example, if we have three spins and we wish to construct states of
total spin \half, we may couple two of the spins into a singlet,
so as to get a doublet:

\beq |0\9\0(|01\9-|10\9)/\sqrt{2}\qquad\mbox{and}\qquad
 |1\9\0(|01\9-|10\9)/\sqrt{2}, \label{doublet}\eeq
where $|0\9$ and $|1\9$ denote the eigenstates of $\sigma_z$, as usual.
The rotations of this doublet generate an irreducible representation
of $SU(2)$.

We can also generate other, equivalent, irreducible representations
by starting with different pairs of spins to make a singlet. These
equivalent representations have of course different physical
meanings. If we used quantum numbers for indicating internal symmetries,
they would have different quantum numbers.

It was shown in \cite{bbbmt,petra} that the use of more than one
equivalent representation does not improve the fidelity of the
transmission. In these articles, the choice of that representation was
irrelevant (of course Alice and Bob had to use the same one). However,
in some cases, that choice may be imposed on us.

As a simple example, consider a given state $|001\9$ of three spins. We
want to split it into a spin~$3\over2$ component and a {\it single\/}
spin \half\ component. This can easily be done, but the spin~\half\
component will not be of the type represented by Eq.~(\theequation),
or by any permutation of the three spins in Eq.~(\theequation). To
see that, we note that spin~$3\over2$ states are symmetric under
permutations of the particles. Therefore we project $|001\9$ on

\beq |\mbox{$3\over2$,$1\over2$}\9 
 =(|001\9+|010\9+|100\9)/\sqrt{3}. \eeq
It follows that the $j={3\over2}$ part of $|001\9$ is 

\beq |\mbox{{$3\over2$},{$1\over2$}}\9
\6\mbox{{$3\over2$},{$1\over2$}}|001\9=(|001\9+|010\9+|100\9)/3. \eeq
What remains of $|001\9$, namely

\beq |001\9-(|001\9+|010\9+|100\9)/3=(2|001\9-|010\9-|100\9)/3,\eeq
is the spin~\half\ part. This is not the direct product of a singlet
and a doublet as in Eq.~(\ref{doublet}). Still that state generates
a perfectly legitimate irreducible representation, with $j={1\over2}$.
(Of course, had we started from a different state, such as $|010\9$, 
we would have ended with a different basis for the irreducible
representation with $j=\half$.)

We have a similar construction, but slightly more complicated, for
Alice's signal having two opposite spins oriented along $z$ and two
others along $x$, as proposed at the end of Sect.~1. It should be clear
that a pair of signals (or any combination of signals) still are {\it
one\/} signal. Alice's signal thus is, with the same notations as above

\beq |A\9=|01\9\0(|0\9+|1\9)\0(|0\9-|1\9)/2=
  (|0100\9+|0110\9-|0101\9-|0111\9)/2. \label{A} \eeq
To find the parts of $|A\9$ with $j=0,1,2$, we proceed as we did for
the case of three spins (we shall not show the explicit calculations,
which are quite lengthy, because they are not necessary for the sequel).

Two points should be emphasized. In earlier works \cite{euler,bbm},
it was not necessary to specify the actual construction of the
irreducible representations that were used. Their choice was arbitrary
and irrelevant. It just had to be the same for Alice and Bob. Now, the
situation is different: these representations are uniquely defined by
Alice's signal in Eq.~(\ref{A}). The choice of that particular signal was
suggested by the Gisin-Popescu method for two spins~\cite{gp}. We
do not claim that it is the optimal signal for {\it two pairs\/} of
spins. To actually investigate optimization, Alice's signal should be
taken as general as possible, namely a sum of 16 terms,

\beq |A\9=a_0|0000\9+\cdots+a_{15}|1111\9, \eeq
and we would have to determine the coefficients $a_n$, subject to
normalization $\sum |a_n|^2=1$. However, this optimization is a long
shot beyond the scope of the present article. Here, we only want to
show that covariant measurements are not always optimal.

\bigskip \newpage

\noindent{\bf4. \ Contravariant quantum measurements}

We now come to Bob's measurement method. As explained in Sect.~1,
Bob examines separately the spins sent by Alice to indicate her $z$
axis, and those indicating her $x$ axis. (The argument below applies
to any number of spins, not just two spins for each axis.) In his
coordinate frame, Bob thus gets two sets of polar angles, $\tz\pz$
and $\tx\px$ respectively, from which he has to infer the Euler angles
$\psi\theta\phi$ that transform Alice's Cartesian frame into his frame.
If Bob's measurements were perfect, the relations between these angles
would be given by equating the Cartesian components of Bob's results
for Alice's $z$ and $x$ axes with the corresponding columns of the
orthogonal transformation matrix~\cite{goldstein}.  This gives

\beq \left( \begin{array}{c}
  \sin\tz\,\cos\pz\\ \sin\tz\,\sin\pz\\ \cos\tz \end{array} \right) 
 = \left( \begin{array}{c}
  \sin\psi\,\sin\theta\\ \cos\psi\,\sin\theta\\ \cos\theta \end{array}
  \right), \label{z}\eeq
and  
\beq \left( \begin{array}{c}
  \sin\tx\,\cos\px\\ \sin\tx\,\sin\px\\ \cos\tx \end{array} \right) 
 = \left( \begin{array}{c}
 \cos\psi\,\cos\phi-\sin\psi\,\cos\theta\,\sin\phi\\
 -\sin\psi\,\cos\phi-\cos\psi\,\cos\theta\,\sin\phi\\
 \sin\theta\,\sin\phi \end{array}
  \right). \label{x}\eeq
These are four independent equations (owing to normalization) for
three unknowns. If Bob's experimental data were exact, a simple solution
would be to obtain from Eq.~(\ref{z})

\beq \theta=\tz\qquad\qquad\mbox{and}\qquad\qquad
 \psi=(\pi/2)-\pz. \label{tp}\eeq
The fidelity of this result, namely for finding the direction of a single
axis by using any number of spins, is discussed in \cite{bbbmt,petra}
where it is shown that, asymtotically, $(1-F)\propto N^{-2}$.

Once $\theta$ is known, $\phi$ can be obtained from the third line of
(\ref{x}):

\beq \sin\phi=\cos\tx/\sin\tz, \eeq
where use was made of the result in Eq.~(\ref{tp}).  Now there is
a difficulty: if $N$ is finite, Bob's estimates are not perfectly
accurate and the right hand side of (\theequation) may be larger
than~1. In general, it is preferable to solve the four equations
(\ref{z}) and (\ref{x}) simultaneously and to seek a best fit for the
three unknowns. The accuracy of this best fit is of course better than
the one given by Eqs.~(\ref{tp}) and (\theequation), where one of the
four original equations was ignored. A simple geometric construction of
the solution is as follows: first, find the direction perpendicular to
the estimated $z$ and $x$ axes; this direction is the best estimate for
the $y$ axis, and therefore for the $zx$ plane. Then, in that plane,
the angle between the estimated $z$ and $x$ axes (given by $\tz\pz$
and $\tx\px$ respectively) is adjusted so that they become exactly
perpendicular. Detailed calculations are under progress (we hope they
will appear in a future publication).

\bigskip 

\noindent{\bf5. \ The dihedral group}

A clearer understanding of contravariant measurements may be gained by
using a finite group. As a concrete example, let us consider
six directions, defined by the polar angles $\theta=45^\circ$
or $135^\circ$, and $\phi=0$ or $\pm120^\circ$. Alice wishes to
indicate one of these directions to Bob. Now, these directions are
the orbit of the dihedral group $D_3$ with six elements: {\bf E}
(the identity), {\bf A}, {\bf B}, {\bf C} (rotations by $180^\circ$
around the symmetry axes $\phi=0$ and $\phi=\pm120^\circ$ in the
$xy$ plane), and {\bf D}, {\bf F} (rotations by $\pm120^\circ$
around the $z$ axis). Here, we are using the same notations as
Wigner~\cite{wigner}. This group has a one-dimensional representation
where all the elements are 1, another where {\bf E}, {\bf D}, {\bf
F} are 1, while {\bf A}, {\bf B}, {\bf C} are $-1$, and there is a
two-dimensional representation, explicitly given in~\cite{wigner}. From
the characters of these three representations, it is possible to find
the contents of any other, reducible one.

Suppose that Alice has a single particle of spin \half, and she wants
to indicate to Bob which one of the six directions she has chosen.
Obviously, she orients her spin along that direction, so that there
are six input states,

\beq \rho=(\1+\mbox{\boldmath$n\cdot\sigma$})/2, \eeq
where $\bn=(\sin\theta\cos\phi,\ \sin\theta\sin\phi, \cos\theta)$.
Likewise Bob has six POVM elements

\beq E_m=(\1+\mbox{\boldmath$m\cdot\sigma$})/6. \eeq
Note that $\sum E_m=\1$. Then the probability of Bob getting result \bm\
is Alice's input is \bn\ is

\beq P(\bm|\bn)={\rm tr}\,(\rho\,E_m)=(1+\mbox{\boldmath$n\cdot m$})/6.
\eeq
Note that the probability of getting the correct result is always 1/3. 

We must now specify a criterion for the fidelity of the transmission. 
A simple one is to give a score~1 is Bob guesses the correct result,
and~0 for all incorrect results. It could also be argued that some
results are more incorrect than others, just as large errors $\chi$ 
in Eq.~(1) are more heavily penalized than small error angles. Here, for
a group of order~6, we could assume that elements belonging to the same
class are less wrong than those belonging to different classes of the
group, and incur a lesser penalty. However, we shall just assume that
all wrong results are equally worthless. Therefore the best that can
be achieved with one spin is fidelity $F=1/3$.

Suppose now that Alice sends several spins.  Rather than using
individual measurements and classical statistics, Bob may perform
joint measurements on all these spins. For example, if there are two
spins, their state belongs to the rotation group representations with
$j=0,1$. However, that group is too rich: we are interested only
in the $D_3$ group which is a subgroup of $SO(3)$. Obviously $j=0$
corresponds to the symmetric one-dimensional representation of $D_3$
(all the elements are~1). As for $j=1$, we have to find the characters
of all the rotations that correspond to elements of $D_3$. This
is very easy, because one may use for $SO(3)$ the real orthogonal
representation, and then, owing to Euler's theorem~\cite{goldstein},
the characters (that is, the traces of the rotation matrices) are equal
to $1+2\cos{\mit \Phi}$, where $\mit\Phi$ is the rotation angle. We thus
find that the character of {\bf E} is 3, those of {\bf A}, {\bf B}
and {\bf C} are $-1$, and those of {\bf D} and {\bf F} vanish. This
means that the triplet state involves the one-dimension representation
with alternating signs and the two-dimensional representation.

Therefore a pair of spin~\half\ particles generates all the
representations of $D_3$, each one once. Taking more spins will not
produce any new irreducible representation. If Bob is restricted to
the use of covariant measurements, the maximal fidelity that can be
achieved is 2/3 (detailed calculations are given in an Appendix).

A simple method which gives better results is the following. Alice sends
$N$ spins, all aligned in the direction she wants to indicate, as in
Ref.~\cite{mp}. This is an angular momentum coherent state~\cite{perel}
for spin $j=N/2$:

\beq  \mbox{\boldmath$n\cdot J$}\,|\psi\9=j\,|\psi\9. \eeq
(This is surely not the optimal strategy. In the present paper, we are
not seeking optimality. We only want to show that some methods give
a better fidelity than a straightforward covariant measurement.) Bob
then performs the covariant measurement for a particle of spin~$j$. His
POVM elements are coherent states as above, with directions uniformly
distributed over the two-dimensional sphere. The overlap of two such
states is~\cite{perel}

\beq \cos^{4j}{(\chi/2)}=\cos^{2N}{(\chi/2)}, \label{expo} \eeq
where $\chi$ is the angle between the true and estimated directions.

Once Bob has found a result $\theta\phi$ (this is what we call the
``raw'' result), he infers (guesses) that the true answer for Alice's
signal is the direction \bn\ closest to $\theta\phi$. It is, as in the
preceding example, the best fit for the answer, knowing the approximate
value given by $\theta$ and $\phi$. As there are {\it finite\/} angles
between the six directions \bn\ that Alice can use, it follows from
(\theequation) that the probability of error decreases {\it
exponentially\/} with $N$. It is plausible that a truly optimal method
would also have such an exponential accuracy, but with a larger
coefficient for $N$ in the exponent.

\bigskip 

\noindent{\bf6. \ Concluding remarks and apologies}

Due to the pressure of a deadline, we did not attempt to find the
optimal strategy in the two examples given above. In both cases, the
same pattern emerges. Bob first executes a POVM which uses a group that
is {\it not\/} the one for Alice's signals. This POVM gives redundant
raw data to Bob, from which he infers, by a classical statistical
analysis, the best estimate for identifying the signal. This final best
estimate is nonlinear and it {\it cannot be obtained directly by a
POVM of rank one\/}, as in Eq.~(\ref{cov}). Indeed, if it could, then
it would have to be a covariant POVM, and we have just shown that this
is not the best method. Explicitly, the POVM elements that give the best
guess are $E_g=\sum E_h$, where the sum runs over all the raw outcomes
$h$ that lead to the same guess $g$.

Note that there is no contradiction with Gleason's
theorem~\cite{gleason} because the proof of the latter refers only
to the outcome of the measuring process, without a further best fit
or other classical statistical analysis. We hope that further research
on this problem will clarify the missing details and give a complete
prescription for the optimal procedure.

\bigskip\noindent{\bf Acknowledgments}

Work by AP was supported by the Gerard Swope Fund and the Fund for
Encouragement of Research. PFS was supported by a grant from the
Technion Graduate School.\newpage

\renewcommand{\theequation}{A \arabic{equation}}
\setcounter{equation}{0}

\noindent{\bf Appendix. \ Dihedral group signals with one or two spins}

In this Appendix we analyze covariant measurements for the detection of
signals belonging to the $D_3$ group. If only one particle of
spin~\half\ is sent by Alice, its state is given as usual by

\beq |2_g\9={\cos(\theta/2)\choose\sin(\theta/2)e^{i\phi}},
 \label{2g}\eeq
where $\theta$ and $\phi$ are the angles that correspond to group element
$g$. Bob's fiducial state is

\beq |B\9=|2_E\9/\sqrt{3}. \eeq
(The factor $\sqrt3$ comes from the order of the group
divided by the number of dimensions~\cite{perel}.) Then

\beq \sum U(g)\,|B\9\6B|\,U^\dagger(g)=\1, \label{Bpovm}\eeq
is Bob's POVM. The fidelity, which is
defined in this problem as the probability of a correct result, is

\beq F=|\62_E|B\9|^2=\mbox{$1\over3$}\,. \eeq

Suppose now that Alice has two spins. She sends them both in state
$|2_g\9$, and Bob tests them separately. The probability that he gets
twice the correct answer is $1\over9$. The probability that he gets the
correct answer once, together with one wrong answer, is $4\over9$. In
the latter case, faced with an ambiguous result, Bob will make a
random choice.  Then the final probability for a correct guess is
${1\over9}+{2\over9}={1\over3}$, exactly as for a single signal!

If Alice sent more than two spins in this way and Bob tested them
separately, the result, given by a multinomial distribution, would
slowly improve. As shown in Sect.~5, Alice and Bob can do much better
by using entangled signals. Let us examine how well they can do if
they have unlimited experimental skill, but are restricted to the use
of covariant measurements.

Alice prepares a signal state within the orbit of group $\cal G$:

\beq |A_g\9=a_0\,|0_g\9+a_1\,|1_g\9+a_2\,|2_g\9, \eeq
where $|0_g\9$ corresponds to the trivial representation (all the
$|0_g\9$ are equal), $|1_g\9$ to the alternating one, and $|2_g\9$
is still given by Eq.~(\ref{2g}). The coefficients $a_m$ are normalized:

\beq |a_0|^2+|a_1|^2+|a_2|^2=1. \label{norm} \eeq
Bob's fiducial vector now is

\beq |B\9=\sqrt{1/6}\,|0_E\9+\sqrt{1/6}\,|1_E\9+\sqrt{1/3}\,|2_E\9,\eeq
so that Eq.~(\ref{Bpovm}) is still valid. 

The probability for a correct result thus is

\beq F=|\6B|A_E\9|^2=|(a_0+a_1)/\sqrt{6}+a_2/\sqrt{3}|^2. \eeq
This is a quadratic expression for the coefficients $a_m$, subject to
the normalization (\ref{norm}). Its maximum is easily found to be
$2\over3$, when

\beq a_0=a_1=1/2\qquad\qquad\mbox{and}\qquad\qquad a_2=1/\sqrt{2}.\eeq
We see that this optimal covariant measurement only improves the
fidelity from $1\over3$ to $2\over3$.  A contravariant measurement
such as the one described in Sect.~4 gives better results: the
infidelity $(1-F)$ decreases exponentially with the number of spins,
owing to Eq.~(\ref{expo}).

\end{document}